%% file: paper.tex
\documentstyle[eqsecnum,aps,epsf]{revtex}

\begin{document}

\title{Solving multi-loop Feynman diagrams using light-front
coordinates}
\author{N.E. Ligterink
\footnote{e-mail: ligterin@ect.it}}
\address{ECT*, Villa Tambosi, Strada delle Tabarelle 286,
I-38050 Villazzano (Trento), Italy}
\date{\today}
\maketitle

\begin{abstract}
We determine the numerical values of scalar multi-loop two-vertex 
Feynman diagrams, the generalized sunset diagrams, by 
integrating  all but the longitudinal momenta
analytically. For the longitudinal momenta we introduce one
collective coordinate, which allows us to determine the numerical
value of the diagram efficiently and to an arbitrary accuracy. 
The imaginary part and the threshold behavior of the
diagram are also handled within this framework.
\end{abstract}


\section{Introduction}

Multi-loop Feynman diagrams pose a serious challenge, especially in
the massive case where multiple scales arise. Apart from the successes 
of high-order expansions in $\alpha$  in QED, which provide a stringent
test of quantum field theory, Feynman diagrams are also the way in which
we understand the most of field theory, since it allows us to
quantify it. They form the basis for our
understanding of many phenomena, such as asymptotic freedom and gauge
invariance.
Therefore, we should strive to refine our handling of Feynman diagrams
and to extend its context and its applications.

In the late 40's, the move from time-ordered perturbation theory \cite{Hei}
towards covariant perturbation theory brought about a revolution
in field theory. \cite{Sch} It created a handle on the calculations and 
people were able the control the divergences. However, despite
the successes for scattering experiments, the method failed to deliver
for bound-states calculations. Therefore,
there has been a constant move "backwards"
to quasi-potential, and time-ordered, formulations, as we can formulate a
bound state only in a single time frame, not with the relative time as it exists
in the covariant formulation. However, often it is hard to follow the route
back from covariant to time-ordered perturbation theory, and then to apply
it to an intrinsic non-perturbative problem, such as a bound-state problem.

One distinct method for describing bound-states in field theory is discrete
light-cone quantization, where as the time direction a light-like direction is
chosen. This has certain advantages, extensively discussed in the literature.
\cite{BP}
However, also here renormalization forms a serious problem, often
dealt with rather callously. Although the renormalization for perturbative
expansions is reasonably under control, \cite{BRS,LBa}
its extension to non-perturbative
calculations is far from desired. 

In this paper I will tackle a class of simple $n$-loop Feynman diagrams,
determine the finite part and show that the light-front approach is 
particularly useful for that. I will also discuss how to look upon 
the large set of counterterms in this highly divergent case.
These diagrams have been studied extensively in the recent years. \cite{Lit}
However, the simplicity of this approach in Minkowski space is striking.
There are no special functions needed, and eventually it depends on
the introduction of the collective coordinate $\beta$, analogous  to
the radial coordinate $r^2$ in Euclidean space. In this case, specifically,
the coordinate interpolates smoothly from the threshold value at the center
of the kinematical to the edges of the kinematical domain.
Eventually, such simple collective coordinates for many-particle systems
might extend the applicability of Hamiltonian light-front field theory,
as, generally, the problem of the Hamiltonian approach is the control 
on number of the variables.

\section{The sunset diagram}

We consider the $n$-loop Feynman diagram, which consists of $n+1$ lines 
between two vertices:
\begin{equation}
{\cal I}_n= \frac{1}{(2 \pi i)^n} \int \frac{d^4k_1 \cdots d^4 k_n}
{ (k_1^2-m_1^2)(k_2^2-m_2^2) \cdots(k_n^2-m_n^2)
((p-k_1-k_2-\cdots -k_n)^2 - m_{n+1}^2)}  \ \ .
\end{equation}
This diagram the generalized sunset diagram. Lines can be added or
removed (see Fig.~\ref{fonion}).
The Feynman diagram is covariant and therefore it is only 
a function of $p^2$ and
the masses $m_1,\cdots, m_{n+1}$. We solve it in the frame where
$p_\perp=(p^1,p^2) = 0$.  We introduce light-front coordinates: 
$k^\pm = \frac{1}{\sqrt{2}}(k^0 \pm k^3)$, and $x_i = k_i^+/p^+$.  
The transverse momenta $k^1$ and $k^2$ are unaltered.
After the residue integration over the  light-front energies $k_i^-$,
we find: \cite{LBb}
\begin{equation}
{\cal I}_n = 
\int_\Delta {\rm d}x_1 \cdots {\rm d} x_n \int {\rm d}^2 k_{\perp 1} 
\cdots {\rm d}^2 k_{\perp n} {1 \over 2^{n+1} x_1 \cdots x_n 
(1-\sum x_i) \left(2 p^+ p^- - \beta^{-1}- \alpha  \right) } \ \ ,
\end{equation}
where
\begin{eqnarray}
\beta^{-1} & = & \frac{ m_1^2 }{ x_1}
\cdots + \frac{m_n^2 }{ x_n} +
\frac{m_{n+1}^2 }{ 1 - \sum x_i} \ \ , \\
\alpha & = &\frac{ k^2_{\perp 1} }{ x_1} +
\cdots + \frac{k^2_{\perp n} }{ x_n} +
\frac{(\sum k_{\perp i})^2 }{ 1 - \sum x_i} \ \ .
\end{eqnarray}
The domain $\Delta$ is given by $x_i \geq 0$ and $\sum x_i \leq 1$.
This integral is the corresponding light-front diagram, equivalent to the 
Feynman diagram.
If we translate the transverse momenta successively starting 
from $k_{\perp n}$:
\begin{equation}
k_{\perp i} = l_{\perp i} - \left( \sum_{j=1}^{j=i-1} k_{\perp j} \right)
\left( \frac{1 }{ x_i} + \frac{1}{ 1 - \sum_{j=1}^i x_j } \right)^{-1} \left( \frac{1}{ 1 - \sum_{j=1}^i x_j } \right),
\end{equation}
we find that the $\alpha $ reduces  to a pure quadratic form in $l_{\perp i}$: 
\begin{equation}
\alpha = \sum_{i=1}^n l_{\perp i}^2 \left({1 \over x_i} + {1\over 1 - \sum_{j=1}^i x_j } \right)\ \  .
\end{equation}
The integral is divergent, and requires counterterms $c_0$, 
$c_1 p^2$, $ \cdots$, $c_{n-1}(p^2)^{n-1}$.
We do so by subtracting the $(n-1)$-th order Taylor expansion in $p^-$ 
around $p^2 =0$ , implemented through the the multiplication
of the integral with the proper moments: \cite{LBa}
\begin{equation}
 {\cal I}_n = \int_\Delta I \to 
{\cal I}_n^{\rm reg} = \int_\Delta I J^n \ \ ,
\end{equation}
where
\begin{equation}
J =  \frac{p^2}{ \beta^{-1} + \alpha }\ \  .
\end{equation}
which will lead to
\begin{equation}
{\cal I}_n^{\rm reg} = \int_\Delta 
\frac{dx_1 \cdots d x_n d^2l_{\perp 1} \cdots 
d^2 l_{\perp n}(p^2)^n }{ 2^{n+1} x_1 \cdots x_n (1 - \sum x_i)
(2 p^+ p^- -\beta^{-1} -\alpha )(\beta^{-1} + \alpha)^n }\ \ .
\end{equation}
If we scale the transverse momenta accordingly and perform the angular 
integrations we obtain the integral:
\begin{equation}
{\cal I}_m^{\rm reg} =
\pi^n \int_0^\infty {d} z_1 \cdots \int_0^\infty {d} z_n \int_\Delta {\rm d} x_1 \cdots
{\rm d} x_n { (p^2)^n  \over 2^{n+1} (  p^2 - \sum z_i - \beta^{-1} )
(\beta^{-1} + \sum z_i)^n } \ \ ,
\end{equation}
where
\begin{equation}
 z_i = l_{\perp i}^2 \left( \frac{1 }{x_i} +
 \frac{1}{ 1 - \sum_{j=1}^i x_j } \right)\ \  .
\end{equation}
In order to integrate over $z_i$ we express the integrand as a series in
$p^2$.
For each separate term we can integrate over all $z_i$,
which yields:
\begin{equation}
{\cal I}_n^{\rm reg} =  -{\pi^n \over 2 ^{n+1}} \int_\Delta {\rm d}^n x 
(p^2)^{n-1}  \sum_{i=0}^\infty {1 \over (n+i)\cdots (i+1)} 
\left( {p^2 \beta } \right)^{i+1}\ \ .
\end{equation}
We can write the series as an analytical function:
\begin{equation}
{\cal I}_n^{\rm reg} 
=\int_\Delta {\rm d}^n x \frac{(-\pi)^n }{2^{n+1} (n-1)! \beta^{n-1}}
 (1 - {\cal T}_{n-1})\left({1 - p^2  \beta } \right)^{n-1} \ln \left[ 
{1 -  p^2 \beta - i \epsilon} \right] \ \ ,
\label{fint}
\end{equation}
where ${\cal T}_{n-1}$ stands for the $(n-1)$-th order 
Taylor expansion around $p^2=0$.
The Taylor expansion yields the following polynomial:
\begin{equation}
{\cal T}_{z (n-1)} (1-z)^{n-1} \ln[1-z] = \sum_{k=1}^{n-1}
 \left(\sum_{j=0}^k -\frac{(-1)^j n!}{(n-j)!j!(k-j+1)} \right) z^k \ \ .
\end{equation}
The imaginary part follows directly from the natural logarithm of
Eq.~(\ref{fint}): 
\begin{equation}
\lim_{\epsilon \to 0} \ln ( -x- i \epsilon)  = \ln |x| - i \pi \theta(x) \ \ ,
\end{equation}
which lead to a finite amplitude, unaffected by the renormalization
procedure which reminiscence appears in Eq.~(\ref{fint}) in the form  of
the subtracted Taylor expansion.
\section{Longitudinal integration}
After the integration over  the $k^-_i$'s and the $k_{\perp i}$'s, we are left
with an integration of the longitudinal momentum fractions $x_i$, which 
cannot be performed analytically. However, note that the integrand of
Eq.~(\ref{fint}) depends 
only on one particular combination of the longitudinal momenta, namely $\beta$.
This $\beta$ is a smooth function of the longitudinal momenta $x_i$, and
ranges between:
\begin{equation}
0 \leq \beta \leq \left( \sum_{i=1}^{n+1} m_i \right)^{-2}  = b \ \ ,
\end{equation}
where for $\beta = b$ the longitudinal momentum fractions $x_i$ equal $m_i \sqrt{b}$.
Therefore  the $n$-dimensional integral over $\Delta$ reduces to
the determination of the integration measure $\mu$ for the integration over
$\beta$:
\begin{equation}
\int_\Delta d^n x f(\beta) = \int_0^b \mu(\beta) d \beta f(\beta) \ \ .
\end{equation}
The volume of the domain $\Delta$ is $\Gamma(n+1)^{-1}$.
Once this measure is determined, it can be used for all values of $p^2$.
The measure $\mu$ can be determined via several means, for example,
with Monte Carlo integration.
The threshold behavior of the diagram is dominated by the 
the values of $\beta $ close to $ b$. For this purpose we can make 
an analytical expansion of $\mu$ around $b$. We find that:
\begin{equation}
\mu(\beta) \sim \frac{1}{2}\Omega_n 
b^{\frac{1-n}{4}}
\left[ \prod_{i=1}^{n+1}\sqrt{m_i}  \right]
(b-\beta)^{\frac{n-2}{2}} \ \ .
\end{equation}
where $\Omega_n$ is the surface area of the unit sphere in $n$ dimensions.
Note that as some of the  masses tend to zero,
the exponent in the measure will be larger than $(n-2)/2$.
The addition of a zero mass particle, $m_i=0$,
leads to a flat direction in $\beta$ with respect to the longitudinal
momentum fraction $x_i$ at the threshold.
Therefore the harmonic approximation breaks down. 
However,
$\beta$ and the measure $\mu$ are well-defined as long as at least 
one particle is massive. 

\subsection{The integration measure}

Apart from series expansion and Monte-Carlo integration mentioned above,
 we can determine the measure
iteratively. Given the integration measure $\mu_n(\beta_n)$ for $n$
momenta, the integration measure $\mu_{n+1}(\beta_{n+1})$
for $n+1$ variables can be expressed  as
\begin{equation}
\int_0^{b_{n+1}} \mu_{n+1}(\beta_{n+1}) d \beta_{n+1}
 = \int_0^{b_n} \int_0^1 {\mu_{n}(\beta_n) }d \beta_n{y^n} d y \ \ ,
\end{equation}
where
\begin{equation}
\beta_{n+1} = \left(\frac{m^2}{1-y} + \frac{1}{\beta_n y} \right)^{-1} \ \ ,
\end{equation}
with $m$ the mass of the added particle, with longitudinal momentum fraction 
$1-y$.
The other longitudinal momenta are scaled by a factor $y$ such that the 
total longitudinal momentum remains 1.

\section{Renormalization}
In this paper we described a method to find the finite part
of any diagram of the type of Fig.~\ref{fonion}.  The divergences are removed
using the Taylor expansion in the external momentum.
The counterterms, $c_0$, $c_1 p^2$ till $c_{n-1} (p^2)^{n-1}$, 
for the sunset
diagram can be expressed as divergent integrals
depending on the masses $m_1, \cdots ,m_{n+1}$: 
\begin{equation}
c_k = \int_\Delta d x_1 \cdots d x_n \int 
\frac{d^2k_{\perp 1} \cdots d^2 k_{\perp n}}{2^{n+1}x_1\cdots x_n (1-\sum x_i)}
\frac{1}{(\alpha + \beta^{-1})^{k+1}} \ \ \ ,
\label{Cs}
\end{equation}
which defines their relation with other renormalization schemes.
The other renormalization schemes will find that the
counterterms in Eq.~(\ref{Cs}) equal an infinite constant, to
be removed, and a finite part, a function of the masses, which is the finite
renormalization.
The use of Taylor expansion became in disfavor, because of two complications.
Firstly, in the case of multiple external momenta, it is not clear which
combination of external momenta should serve as variable in the Taylor
expansion; different choices will lead to different results, and do
not automatically guarantee locality. Secondly, in the case of gauge
theories, an extremely consistent scheme, which treats a whole class of
integrals in the same way, is required such that the 
gauge invariance is preserved. Dimensional regularization has for
a long time been the only scheme satisfying this consistency, which
preserved algebraic relations existing among different integrands 
of Feynman integrals. For example, the fermion-loop correction to
the gauge propagator $\Pi^{\mu\nu}$ must be transverse, therefore
the two parts to $\Pi^{\mu\nu}$, namely $g^{\mu\nu} \Pi_s$ and 
$p^\mu p^\nu \Pi_t$  must be handled in the same way such that
$p^2 \Pi_t = -\Pi_s$, which is difficult problem for an arbitrary 
regularization scheme, since the two terms have different degrees
of divergence.

However, for an Hamiltonian approach, such as light-front field
theory, the renormalization at the level of the integrand is required,
if one wants to carry the renormalization procedure 
over from the covariant renormalization.
Note that a straightforward cut-off procedure breaks covariance, since
it cannot be applied to the energy part of the covariant integration.
The integration of the energies and the regularization should
be interchangeable, such that locality is guaranteed. So, although it leads
to further complications which requires careful analysis, the Taylor
expansion is the way forward for the Hamiltonian approach. \cite{SBK}

The natural choice of renormalization for a Hamiltonian approach, is
to make the self-energy contributions vanish as all the particles 
are on-shell. However, this is not consistent with the covariant, local
and therefore true, renormalization. If the sum energy is the sum of
the on-shell energies, it does not mean that the energy is shared
out evenly; a large part of the amplitude might arise from 
the case that both particles
are off-shell in different directions. Therefore it is essential
to treat the subtractions as pure constants.

Even more, although we can generate finite terms in light-front perturbation
theory, for a proper light-front approach we should take the procedure
one step further and determine the corresponding finite wave function.
However, this is far beyond the scope of this paper.
 
\section{Results}
Central to this approach is the actual shape of the measure $\mu(\beta)$.
For particles of equal mass, we find that the measure is most spread
over the whole range of $\beta$.
As the masses start to deviate the measure peaks more and more at low
values of $\beta$. However, the measure stays finite, even for massless
particles. In Fig.~\ref{fsym} we show two scaled, normalized set of measures, one
for equal mass particles, and one for particles with increasing masses.
Note that the increasing masses peak more at lower values of $\beta$, 
due to the leading contributions from the heavy particles carrying 
large momentum fractions. In Fig.~\ref{fasym} we compare the measures for
two massive particles and a number of particles with equal, but small, 
or vanishing, masses.

For the inspection of the amplitudes, I fitted the measures with 
a five parameter function, which fits
the measure with an accuracy within a few percent. The accuracy for
the massless case is higher than for the massive case, in the former case
it is below a percent.
The function depend on the parameters $\gamma_1,\gamma_2,\cdots,\gamma_5$:
\begin{equation}
\bar \mu (\bar \beta) = \gamma_1(1 - \bar \beta )^{\gamma_2} + \gamma_3(1 - \bar \beta)^{\gamma_5}\bar 
\beta^{\gamma_4} \ \ ,
\end{equation}
where $\bar \beta = \beta/b$ and $\bar \mu = \Gamma(n+1) \mu/b$, such that 
the axis and the measure are normalized to unity. The fitted 
parameters for the two extreme cases,
one case with all masses equal, and one case with the first two masses 
1.0, and all the other masses zero are given respectively in Table~\ref{tsym}
and Table~\ref{tasym}.

\section{Conclusions}

I have derived a straightforward, and largely analytical, way
to determine the finite part of the sunset diagram. Both the
threshold behavior and the full amplitude can be determined 
accurately. I removed the divergent parts by subtracting the 
corresponding Taylor expansion. The mass dependence of the diagram,
for the scaled external momentum $b p^2$, is only weak over the whole
range of masses, and this dependence appears solely
in the integration measure $\mu(\beta)$.

A careful analysis of the Feynman parameterization \cite{Lit} could also yield a
similar variable $\beta$. However, it requires one to work in Euclidean
space, where the imaginary part does not come  for free.
Also the poles in dimension space, which are the different 
subdivergencies of the integral, are transferred to singularities in
the parameter space, which renders dimensional regularization invalid.
Therefore the removal of the lower-order Taylor expansion is essential.

It seems possible to extend the light-front approach to more 
complicated diagrams,
which could contain two and more light-front intermediate states. 
This requires the introduction of more $\beta$ variables. The determination
of the measures stays essentially the same.  Eventually, the light-front
approach might be more convenient for the calculation of multi-loop 
diagrams, as it sees
such a diagram as a transition via collection of successive intermediate 
states, which can all be handled separately, and do not grow as wildly as
the number of the subgraphs of a complicated covariant multi-loop
Feynman diagram.

\input{figures.tex}
%
%

\begin{table}[h]
\caption{The fitted parameters for the equal mass case.}
\begin{center}
\begin{tabular}{c|ccccc}
    $m$ & $\gamma_1$ & $\gamma_2$ & $\gamma_3$ & $\gamma_4$ & $\gamma_5$ \\ \hline
$1.0,1.0,1.0$& .661422  &  1.01707  & 1.23111 & .815085 &  .0064803  \\
$(1.0)^4$ & .746938  & .476909 & 1.66776 & .814929 & .526326  \\
$(1.0)^5$ &  .817047  &  1.08804  &  3.45241 &  .950327 & .982281 \\
$(1.0)^6$ &  .874748   &  1.55609  &   5.90012 &  1.06571 &  1.44554  \\
$(1.0)^7$ &  .921155   &  1.94665 & 9.09230  &  1.16310  &  1.88020
  \end{tabular}
 \end{center}
\label{tsym}
\end{table}

\begin{table}[h]
\caption{The fitted parameters for the massless case.}
\begin{center}
\begin{tabular}{c|ccccc}
$m$ &    $\gamma_1$ & $\gamma_2$ & $\gamma_3$ & $\gamma_4$ & $\gamma_5$ \\ \hline
$ 1.0,1.0,0.0$& .988223   &   .538897  &   1.07238  &   .706459 &    .490497  \\
$(1.0)^2,(0.0)^2$ &  1.48154    &  1.33267  &   2.60100  &   .706458 &    1.66423   \\
$(1.0)^2,(0.0)^3$ &  1.97032  &   2.19627   &  4.63872   &  .702103  &   2.75778  \\
$(1.0)^2,(0.0)^4$ &  2.45665  &   3.05772   &  7.09066   &  .698004  &   3.84503   \\
$(1.0)^2,(0.0)^5$ &  2.94061  &   3.92001   &  9.91789   &  .694266  &   4.92679 
  \end{tabular}
\label{tasym}
 \end{center}
\end{table}
\end{document}

%% file: figures.tex
\begin{figure}[h]
   \epsfxsize=9cm
   \centerline{\epsffile{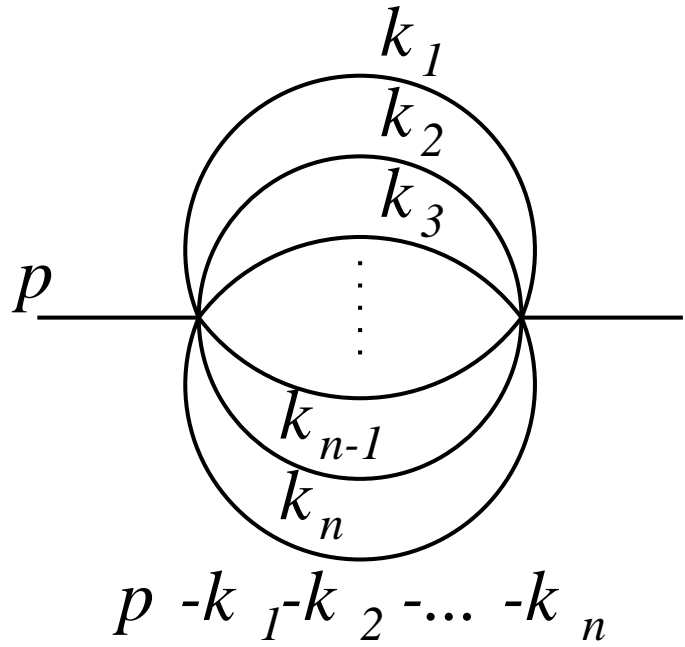}}
   \caption{The generalized sunset diagram}
\label{fonion}
\end{figure}

\begin{figure}[h]
   \epsfxsize=9cm
   \centerline{\epsffile{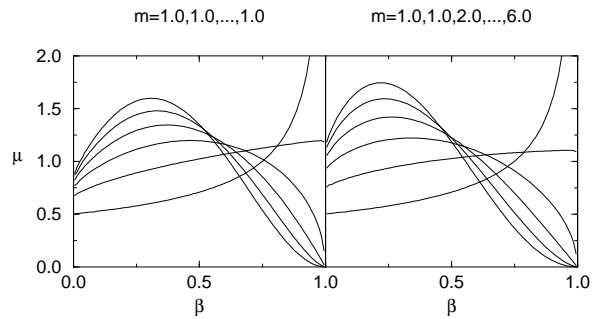}}
   \caption{Two sets of integration measures, all normalized to unity,
left with one to six loops, where all the masses are equal, and
right, with one to six loops including heavier and heavier particles:
$\{1.0,1.0\}$, $\{1.0,1.0,2.0\}$, $\{1.0,1.0,2.0,3.0\}$, $\cdots$,
$\{1.0,1.0,2.0,\cdots,6.0\}$. The variable $\beta$ is scaled such that it
runs from zero to one.}
\label{fsym}
\end{figure}

\begin{figure}[h]
   \epsfxsize=9cm
   \centerline{\epsffile{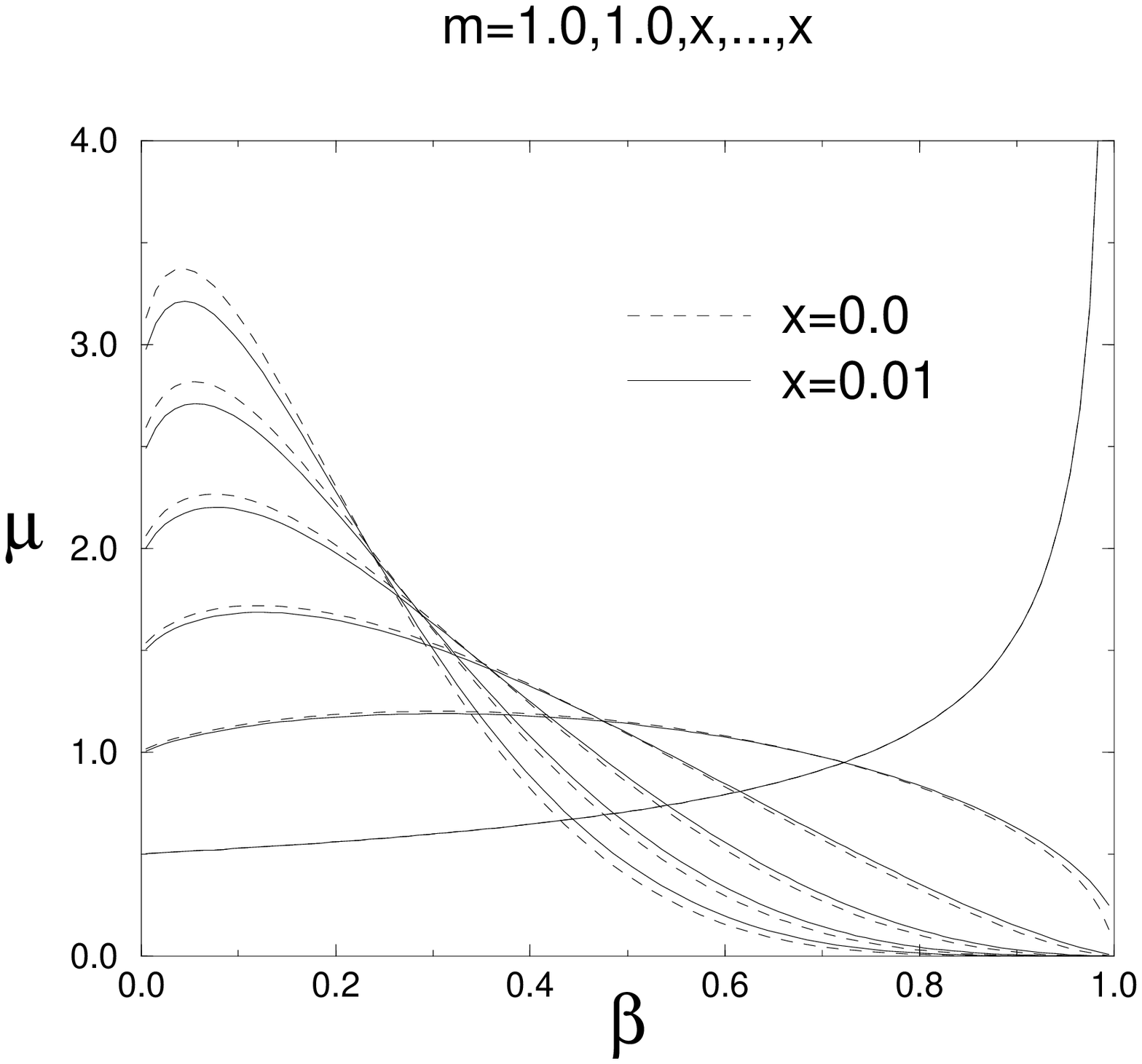}}
   \caption{Two sets of integration measures, all normalized to unity,
left with one to six loops, where the first two masses are 1.0, and the
other masses are small (0.01) for the solid line, or zero, for the dashed line.
Note the rapid decline at $\beta=1$.}
\label{fasym}
\end{figure}

%% file: paper.bbl
\begin{thebibliography}{100}
\bibitem{Hei} W. Heitler, {\it The quantum theory of radiation},
          (Clarendon Press, Oxford, 1954).
\bibitem{Sch}J. Schwinger, {\it Selected papers on quantum electrodynamics},
          (Dover, New York, 1958).
\bibitem{BP}S.J. Brodsky and H.C. Pauli, in {\it the 30th Schladming Winter 
   School; Recent aspects of quantum fields},
 (Spinger Verlag, Berlin, 1991), and the references therein.
\bibitem{BRS}S.J. Brodsky, R. Roskies, and R. Suaya,
       Phys. Rev. D {\bf 8}, 4574 (1973).
\bibitem{LBa}N.E. Ligterink and B.L.G. Bakker,
        Phys. Rev. D {\bf 52}, 5917 (1995).
\bibitem{Lit}A.I. Davydychev and V.A. Smirnov,
      Nucl. Phys. {\bf B554}, 391 (1999);
      S. Groote, J.G. K\"orner, and A. A. Pivovarov,
      Nucl. Phys. {\bf B542},  515 (1999);
      S. Groote, J.G. K\"orner,  and A. A. Pivovarov,
     Phys. Lett. {\bf 443 B}, 269 (1998);
     F.A. Berends, A.I. Davydychev, and  N.I. Ussyukina
     Phys. Lett. {\bf B426}, 95 (1998);
     F.A. Lunev,
     Phys. Rev. D {\bf 50}, 7735 (1994).
\bibitem{LBb}N.E. Ligterink and B.L.G. Bakker,
         Phys. Rev. D {\bf 52}, 5954 (1995).
\bibitem{SBK}N.C.J. Schoonderwoerd and B.L.G. Bakker, 
       Phys. Rev. D {\bf 58}, 025013 (1998);
       N.C.J. Schoonderwoerd, B.L.G. Bakker, and V.A. Karmanov, 
       Phys. Rev. C {\bf 58}, 3093 (1998).
\end{thebibliography}
